# Photoabsorption by Volume Plasmons in Metal Nanoclusters

Chunlei Xia, Chunrong Yin, and Vitaly V. Kresin

*Department of Physics and Astronomy*
*University of Southern California*
*Los Angeles, CA 90089-0484*

**Abstract**

It is well known that plasmons in bulk metals cannot be excited by direct photoabsorption, that is, by coupling of volume plasmons to light. Here we demonstrate that the situation in nanoclusters of the same metals is entirely different. We have carried out a photodepletion measurement for $Na_{20}$ and $Na_{92}$ and identified a broad volume plasmon absorption peak centered slightly above 4 eV, revealing the possibility of optical excitation of volume-type collective electronic modes in a metallic system. The observed phenomenon is related to different selection rules for finite systems.

*1.Introduction*. Collective oscillations dominate the response properties of finite quantum systems such as nanoparticles, fullerenes, and nuclei. These normal modes of delocalized electron or nucleon clouds share the property that their spectra reflect the microscopic coherence resulting from particle-particle interactions [1]. In an infinite solid (e.g., metal or graphite) the description of collective motion is simplified by translational invariance, which imposes a plane-wave character on the oscillations and enforces strict selection rules on their excitation. For example, volume plasmons in bulk metals cannot be excited optically: light waves are transverse, whereas the compressional plasma waves in an infinite medium are longitudinal [2]. Therefore optical spectroscopy could not be brought to bear on these important elementary excitations, and they have been primarily probed by electron energy-loss spectroscopy [3]. The situation becomes very different in nanoparticles, even those of the same metal. Indeed, when a boundary surface is present, the wave vector no longer remains a good quantum number and the photoabsorption spectrum becomes richer, making it possible to access a greater variety of collective modes by optical spectroscopy. As a result, in finite metal nanoclusters the compressional volume plasmons are expected to become capable of coupling to light. This work provides the first experimental confirmation of this effect in free size-selected metal clusters.

Dipolar vibrations of delocalized valence electrons in nanoclusters of simple *s*-electron metals, such as the alkalis, have been extensively investigated both experimentally and theoretically (see, e.g., the reviews [4-7] and references therein) since their first observation twenty years ago [8]. They manifest themselves as intense photoabsorption peaks in the visible part of the spectrum and are commonly referred to as *Mie, or surface, plasma resonances* [9] because for larger particles they correspond to translational electron oscillations leading to surface-localized charge modes [6,10,11]. However, the adjacent ultraviolet part of the cluster photoabsorption spectrum has remained practically unexplored experimentally, despite theoretical predictions that an appreciable fraction of the dipole oscillator strength of the cluster valence electrons should reside there (e.g., [4,6,12-17]). It is this photoabsorption region that has been established by calculations to derive from compressional electron oscillations, or *volume plasma resonances*, red-shifted from the bulk plasmon frequency $\omega_p$ [4,6,11,13]. Fig. 1 illustrates this predicted behavior. Of course, in small systems the collective excitations can hybridize and lose their pure "surface" and "volume" character, but for convenience the high- and low-frequency excitations continue to be referred to by their macroscopic labels.

As indicated above, an analogous situation exists in nuclear physics, where the giant nuclear dipole resonance is represented by a combination of "Goldhaber-Teller" and "Steinwedel-Jensen-Migdal" modes, corresponding to surface-like and volume-like oscillations, respectively. A similar effect was recently demonstrated for the $C_{60}$ fullerenes: resonance-enhanced ionization by synchrotron radiation revealed a surface [18] and a volume [19] collective mode, both of which are dipole active [20-22,11].

Motivated by these considerations, we have carried out a measurement of photoabsorption cross sections of a pair of prototypical metal nanoclusters $Na_{20}$ and $Na_{92}$, covering the near-UV, as well as the visible, parts of their resonance spectra [23].

*2.Experiment.* The measurements were performed in the now-standard longitudinal-beam-depletion geometry [27] first applied to cluster beams in [28,8]. An outline of the arrangement is as follows [29]. A supersonic beam of neutral sodium clusters was generated by co-expanding metal vapor and argon gas through a heated nozzle [5]. At the end of the 2.1 m-long free flight path, the particles were ionized by filtered ultraviolet light from an arc lamp and focused into a quadrupole mass spectrometer set to transmit the cluster size of interest.

The beam was illuminated by 3-6 ns counterpropagating laser pulses, produced by a tunable Nd:YAG/OPO-based laser system and steered by two prisms into the end window of the detector chamber located 1.5 m from the laser. Upon absorbing a photon, clusters rapidly evaporate, so that following a laser pulse the detector counting rate of the selected cluster size drops and remains depleted for ≈2 ms (corresponding to the beam flight time from the skimmer to the detector entrance). This "depletion profile" was recorded by a multichannel scaler synchronized with the laser. The ratio $r$ of the depleted to the equilibrium counting rates, and the photon flux $\phi$ yield the photoabsorption cross section $\sigma$: $r=\exp(-\sigma\phi)$ for one-photon absorption. The photon flux was monitored by inserting a laser pulse energy meter in the middle of the beam region used for extracting $r$, before and after acquiring each depletion profile (~2-4 min long, i.e., ~2000-4000 laser pulses) [30]. In order to ensure that the cluster beam was overlapped by the central portion of the laser beam profile, three collimating apertures were spaced along the flight path, with the aforementioned energy meter located near the middle aperture, and a calibrated photodiode near the upstream one. The facts that their two readings remained proportional to each other and that the value of the ratio $r$ remained consistent for several different channel ranges of the depletion profile, testify to a good degree of alignment. Data points were acquired in 5 nm intervals between 220-420 nm and 10 nm intervals between 420-600 nm.

*3.Results and discussion.* The measured photoabsorption cross sections of $Na_{20}$ and $Na_{92}$ are presented in Figs. 2 and 3, respectively. The average estimated uncertainties in the cross section values are 15% for $Na_{20}$ and 20% for $Na_{92}$.

The visible portions of both plots exhibit the well-known strong surface-plasmon resonance, fragmented into discrete peaks (see below and cf. Fig. 1). As seen in Fig. 2(A), its position, shape and strength for $Na_{20}$ are in very good agreement with the spectra published previously [31,32]. (To the best of our knowledge, there have been no prior measurements for $Na_{92}$.) An important new part of the data is the considerable tail extending into the UV.

Before proceeding with a quantitative discussion, it may be worth noting that both clusters display a weak undulation between 3.5 and 4.0 eV, which is close to their ionization potentials of 3.75 eV ($Na_{20}$) and 3.47 eV ($Na_{92}$) [33,34] and is therefore possibly related to the opening of the continuum channel for electron emission [35].

For a more detailed analysis of the spectra we fit them by Lorentzian line shapes (cf., e.g., [31,32,36,37]). The $Na_{92}$ spectrum was fitted to three peaks, and for $Na_{20}$ either three or four peaks could be used, the latter choice appearing more accurate [38]. The fitted peaks are depicted in Figs. 2(B) and 3, and their parameters are listed in Table 1. The fragmentation of the photoabsorption strength has been explained as the result of a coupling between collective motion and individual electronic transitions (see, e.g., [4,6,14,16] and references therein). This may be viewed as the size-quantized limit of Landau damping of plasma waves, or equivalently as a reflection of the fact that the collective modes are envelopes of coherent superpositions of discrete electronic transitions [39,40].

Of particular interest for our present purposes is the broad peak slightly above 4 eV, labeled Peak 1 in Table 1 and shaded in the figures. In $Na_{20}$ it is slightly more red-shifted ($\omega=0.68\omega_p$, where $\omega_p=5.9$ eV is the bulk free-electron sodium plasma frequency) and comprises a higher fraction of the area of the listed peaks (19%) than in $Na_{92}$ ($0.71\omega_p$ and 14%). This behavior is in clear correspondence with that expected for a dipolar "volume plasmon" resonance in Fig. 1. The absolute magnitudes of the frequency and the red shift are also in very good agreement with theoretical predictions for such a resonance: e.g., the analytical calculation in [4] yielded $\approx 0.75\omega_p$ for its position, and $\approx 20\%$ and $\approx 15\%$ for its weight in $Na_{20}$ and $Na_{92}$, respectively. The manifest match between the behavior and parameters of the UV feature and the predicted nanocluster "volume plasmon" resonance enables us to identify them with each other.

Moreover, measurements in the 230-350 nm region for $Na_{19,21,57}$ attested that these clusters also have similar photoabsorption cross sections per atom in this range [41].

It should be remarked that the peaks in Figs. 2 and 3 do not encompass the full valence electrons' dipole oscillator strength $f$ [42]: the peaks in Table 1 correspond to $f=71\%$ for $Na_{20}$ and $f=60\%$ for $Na_{92}$. Therefore additional photoabsorption channels must be present elsewhere. For one, the excitation of individual valence electrons into the continuum has been predicted to produce a low but very long photoionization tail [43], thereby accommodating an appreciable amount of these electrons' oscillator strength. Furthermore, a constraint imposed by the clusters' static electric polarizability implies that in addition to these putative high-frequency transitions there must also be additional ones at low frequencies [44]. Thus the high energy ionization tail would need to be accompanied by enhanced valence electron absorption in the IR (cf. [47]), most likely due to individual electron-hole excitations. A search for cluster electron photoabsorption at both of these frequency limits, while experimentally quite challenging, would be very interesting and informative.

In summary, the photoabsorption spectra reported here provide the first experimental observation of optically excited "volume plasmon" collective electronic states in metal nanocluster particles, a phenomenon unique to finite systems.

We would like to thank R. Rabinovitch for valuable experimental help and discussions, R. Moro for very useful input on data analysis, and L. Hand for support with laser operation. This research was supported by the U.S. National Science Foundation (PHY-0652534).

**Table 1.** Peak energies ($E$), widths ($\Gamma$, FWHM), and areas ($A$, per atom) derived from fits to the cluster photoabsorption profiles. Quantities in parentheses are estimated uncertainties in the fitting parameters.

**Na$_{20}$**

|  | $E$ (eV) | $\Gamma$ (eV) | $A$ (eV·Å$^2$) |
|---|---|---|---|
| Peak 1 | 4.04 *(0.03)* | 1.19 *(0.07)* | 0.15 *(0.01)* |
| Peak 2 | 3.17 *(0.02)* | 0.52 *(0.02)* | 0.13 *(0.01)* |
| Peak 3 | 2.77 *(0.01)* | 0.23 *(0.01)* | 0.10 *(0.01)* |
| Peak 4 | 2.42 *(0.01)* | 0.26 *(0.01)* | 0.40 *(0.01)* |

**Na$_{92}$**

| Peak 1 | 4.20 *(0.08)* | 1.16 *(0.15)* | 0.093 *(0.016)* |
|---|---|---|---|
| Peak 2 | 3.24 *(0.03)* | 0.37 *(0.04)* | 0.07 *(0.01)* |
| Peak 3 | 2.80 *(0.01)* | 0.42 *(0.01)* | 0.50 *(0.02)* |

**Figure Captions**

**Fig. 1**. A depiction of the theoretically predicted evolution of the photoabsorption strength $S$ of "surface" and "volume" plasmons in nanoclusters as a function of size (adapted from [6,11]). Microscopically, these collective modes are envelopes of coherent superpositions of electron-hole excitations, and because of the discreteness of electronic states in clusters they are split among several transitions and broadened, as indicated by the dashed lines. The "surface" and "volume" labels derive from the fact that the two resonance regions can be shown to correspond to two different types of electron density oscillations - translation vs. compression - whose dipolar components are schematically outlined by the diagrams at the top (the gray rectangle represents the positive charge distribution, and the dotted lines the oscillating valence electron density). In finite particles both collective resonances are red-shifted with respect to the bulk frequencies (labeled $\omega_{Mie}$ and $\omega_p$, respectively), and both can couple to light, whereas in the bulk limit the volume plasmon cannot.

**Fig. 2.** (A) Photoabsorption cross sections per atom of $Na_{20}$ nanoclusters. The thick line is drawn through the individual experimental data points that are shown as diamonds in (B), the squares and circles are data from [31] and [32], respectively.

(B) Diamonds: photoabsorption cross sections of $Na_{20}$; solid line: fit to the data by means of a sum of Lorentzian profiles (dashed lines). The shaded peak is the volume plasmon which becomes dipole-active in finite particles.

**Fig. 3**. Photoabsorption cross sections and fits for $Na_{92}$, marked as in Fig. 2(B).

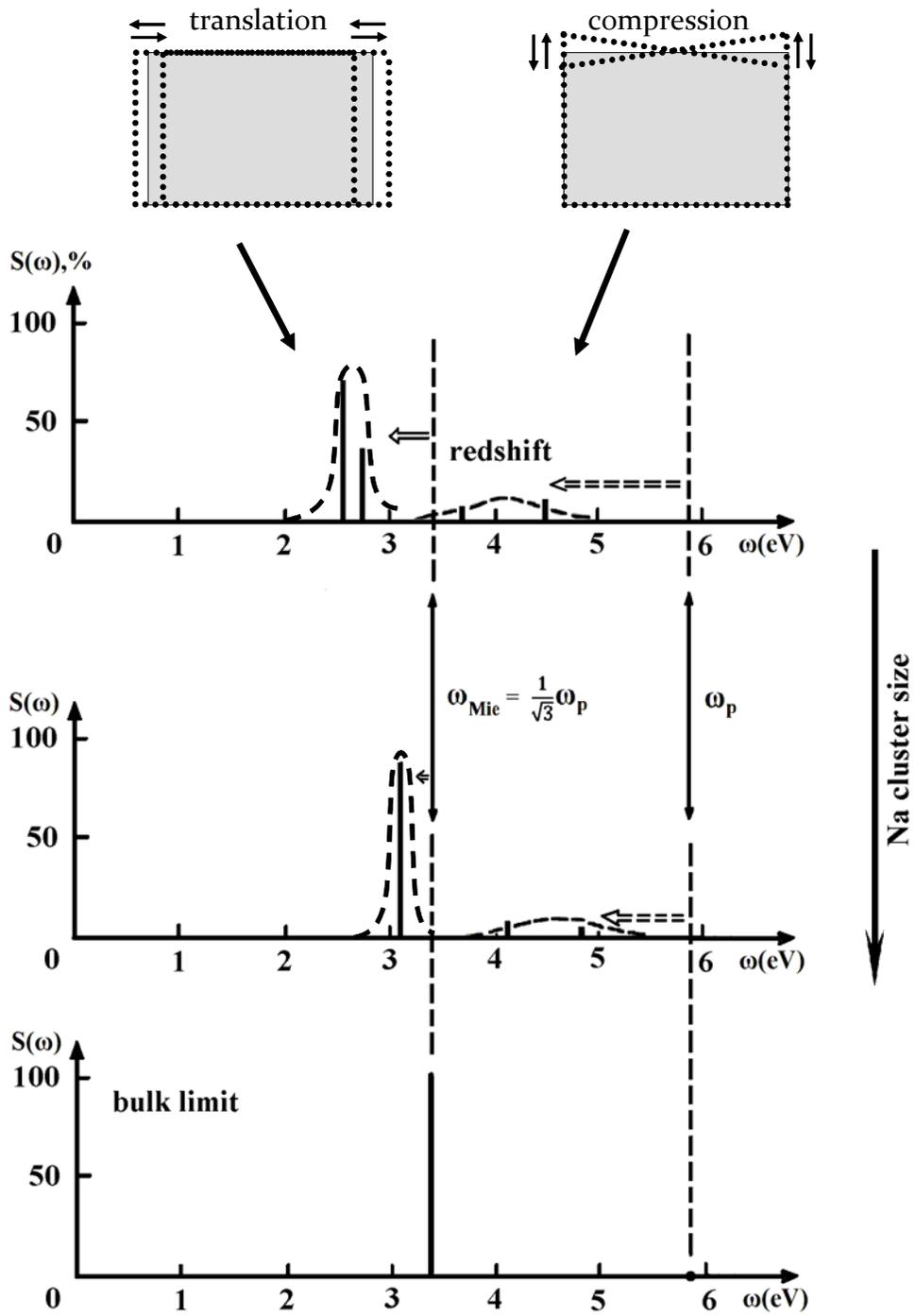

**Fig. 1**

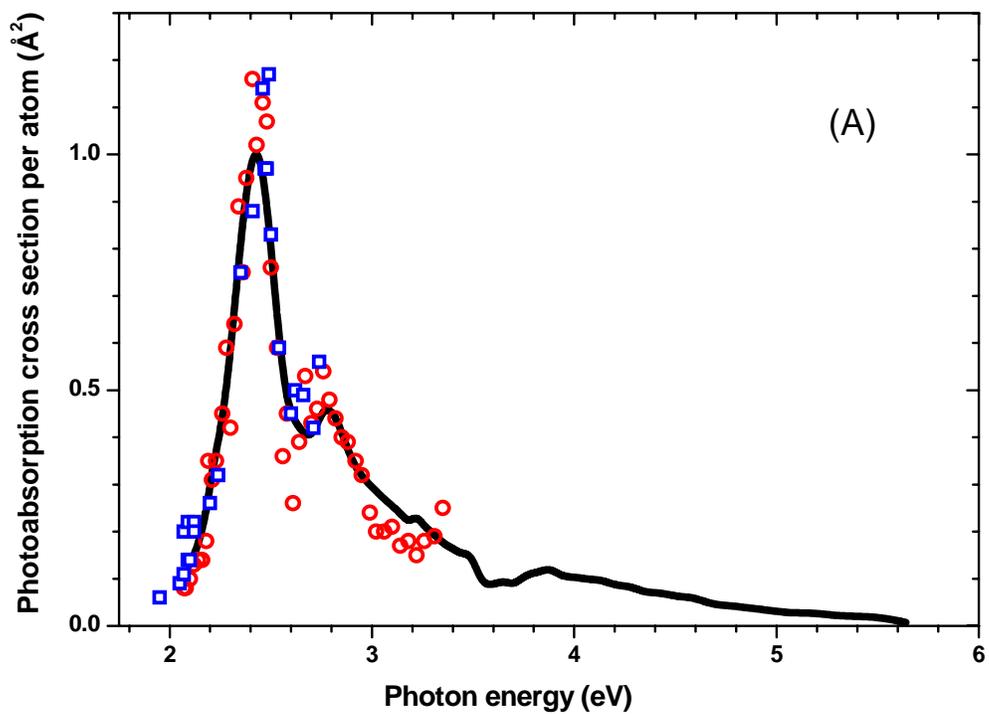

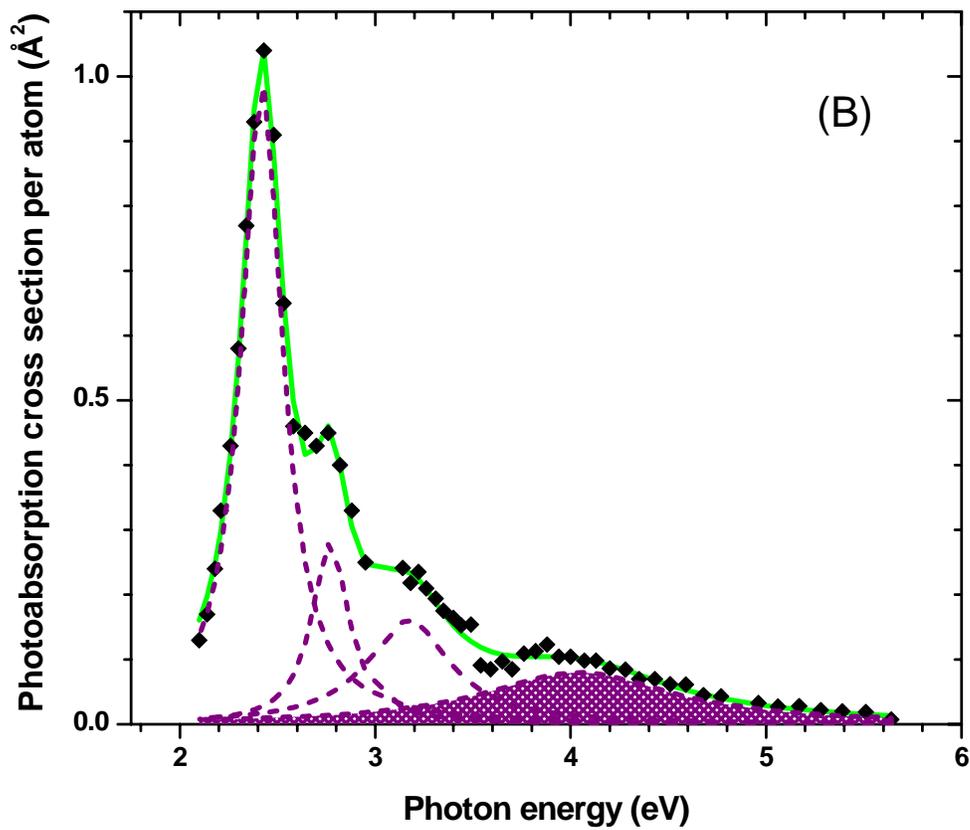

**Fig. 2**

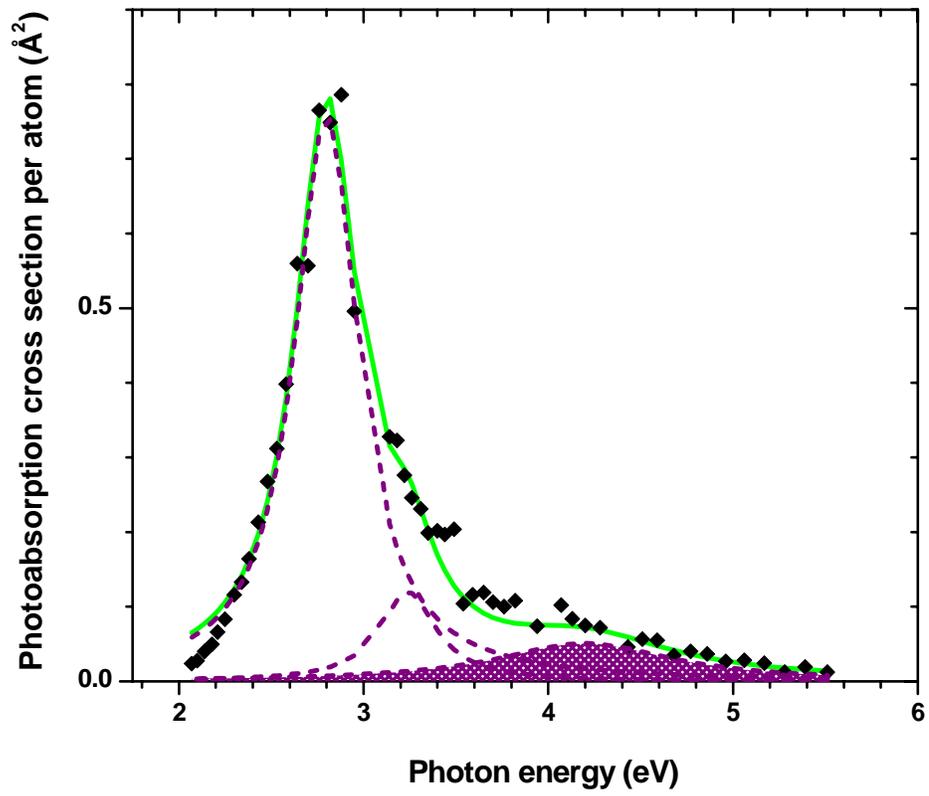

**Fig. 3**